\documentclass[conference]{IEEEtran}
\IEEEoverridecommandlockouts

\usepackage{cite}
\usepackage{amsmath,amssymb,amsfonts}
\usepackage{algorithmic}
\usepackage{graphicx}
\usepackage{textcomp}
\usepackage{xcolor}
\usepackage{bm}
\usepackage{booktabs}
\usepackage[hidelinks]{hyperref}
\def\BibTeX{{\rm B\kern-.05em{\sc i\kern-.025em b}\kern-.08em
    T\kern-.1667em\lower.7ex\hbox{E}\kern-.125emX}}
\begin{document}

\title{DiSR-ECG: Residual Shifting Conditional Diffusion for Robust ECG Super-Resolution\\
{\footnotesize }
}

\author{\IEEEauthorblockN{Fan-Yi Hsu\IEEEauthorrefmark{1}\IEEEauthorrefmark{2},
I Chiu\IEEEauthorrefmark{2}, Kuan-Chen Wang\IEEEauthorrefmark{2}, Sheng-Yu Peng\IEEEauthorrefmark{1}, and
Yu Tsao\IEEEauthorrefmark{2}}
\IEEEauthorblockA{
\IEEEauthorrefmark{1}National Yang Ming Chiao Tung University, Hsinchu, Taiwan\\
\IEEEauthorrefmark{2}Academia Sinica, Taipei, Taiwan \\
Email: fanyi194.cs12@nycu.edu.tw,\ 
d13949002@ntu.edu.tw,\
d12942016@ntu.edu.tw,\\
speng@nycu.edu.tw,\
yu.tsao@citi.sinica.edu.tw
}
}
\maketitle
\begin{abstract}
Electrocardiogram (ECG) signals are essential for arrhythmia diagnosis. With the growing adoption of long-term monitoring via wearable and portable devices, energy-efficient acquisition has become critical, motivating the development of ECG super-resolution (SR) techniques to reconstruct high-resolution signals from low-sampling-rate inputs. While recent SR methods based on discriminative neural networks have shown strong performance, their robustness under distribution shift remains uncertain, which poses a key challenge for real-world deployment. In this study, we propose DiSR-ECG, a residual shifting conditional diffusion framework for robust ECG SR. The model integrates residual shifting with Mamba-based temporal modeling and conditional guidance to enable accurate multi-lead reconstruction. Experiments on two large-scale ECG databases—PTB-XL for in-domain evaluation and Chapman-Shaoxing for cross-dataset evaluation—demonstrate that DiSR-ECG achieves state-of-the-art performance in both settings, with particularly strong generalization under distribution shift. These results highlight the potential of DiSR-ECG to provide reliable and clinically applicable ECG reconstruction for real-world monitoring.
\end{abstract}

\begin{IEEEkeywords}
Electrocardiography, deep learning, diffusion,  super-resolution, wearable device
\end{IEEEkeywords}

\section{Introduction}

Cardiovascular diseases are the leading cause of death worldwide, with sudden cardiac death (SCD) from arrhythmias such as ventricular tachycardia and fibrillation posing a major public health concern~\cite{KAPLANBERKAYA2018216}. Electrocardiogram (ECG) signals are a non-invasive tool for monitoring heart activity, making them essential for arrhythmia diagnosis and long-term patient monitoring~\cite{hurst1998naming}. With the widespread adoption of wearable and portable devices for continuous monitoring, power-efficient data acquisition has become increasingly critical~\cite{huda2020low,xia2018automatic}. A practical strategy to reduce power consumption is to record ECG signals at lower sampling rates~\cite{nishikawa2018sampling} and apply super-resolution (SR) techniques to reconstruct high-resolution (HR) data from their low-resolution (LR) counterparts. However, conventional SR methods, such as linear or cubic spline interpolation, often introduce artifacts that distort ECG morphology and compromise diagnostic reliability~\cite{pizzuti1985digital,krylov2009combined}, limiting their clinical applicability.

To address these challenges, neural networks (NNs) have been widely adopted for ECG SR due to their strong nonlinear modeling capacity and data-driven nature. For example, SRECG~\cite{chen2023srecg} adapted the SRResNet~\cite{ledig2017photo} architecture from image processing for ECG enhancement. By jointly optimizing regression and classification losses, SRECG partially restored the diagnostic utility of LR signals in arrhythmia classification. Autoencoder-based frameworks, such as DCAE-SR~\cite{lomoio2024dcae}, further explored convolutional encoder--decoder designs to capture fine-grained waveform details. While effective, these purely convolutional methods are limited in modeling long-range temporal dependencies. More recently, MSECG~\cite{lin2025msecg} combined convolutional layers with the Mamba state-space model, enabling simultaneous learning of local waveform structure and broader temporal context. This design achieved state-of-the-art performance on the PTB-XL database~\cite{wagner2020ptb} under noisy conditions. However, the robustness of these discriminative SR methods under distribution shifts remains uncertain, which may pose a critical challenge for real-world deployment. This limitation motivates the exploration of generative approaches, which explicitly model data distributions and have shown stronger robustness under unseen conditions~\cite{NEURIPS2022_95504595,lu2022conditional}.

\begin{figure*}[t]
    \centering    \includegraphics[width=1\linewidth]{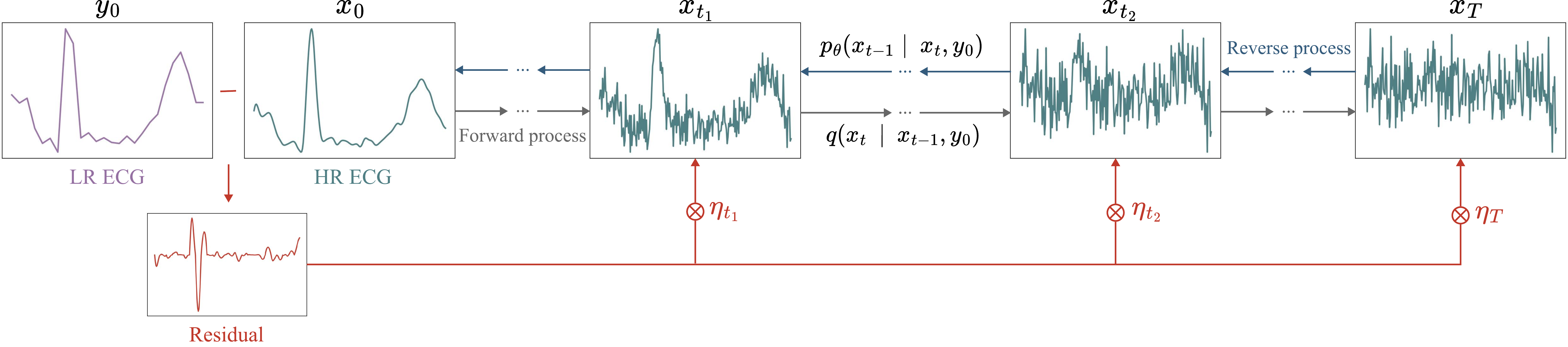}
    \caption{Overview of residual shifting diffusion.}
    \label{fig:diffusionprocess}
\end{figure*}

Diffusion models have emerged as a powerful generative framework for data synthesis and restoration. Compared to variational autoencoders (VAEs) and generative adversarial networks (GANs), diffusion models provide more stable training and greater capacity to capture complex distributions~\cite{ho2020denoising, nichol2021improved}. By iteratively denoising noisy inputs, they have exhibited strong results in generating high-fidelity signals, including audio and biomedical signals such as ECG and surface electromyography~\cite{kong2021diffwave,li2023descod,liu2024sdemg}. Moreover, several studies suggest that diffusion-based methods can generalize better under domain mismatch~\cite{lu2022conditional, NEURIPS2022_95504595,li2024biodiffusion}, highlighting their potential for developing robust ECG SR under distribution shift.

In this study, we propose DiSR-ECG, a residual shifting conditional diffusion framework for ECG SR. The model integrates residual shifting with Mamba-based temporal modeling and conditional guidance from LR inputs to enable accurate multi-lead reconstruction. We conduct experiments on two large-scale ECG databases: PTB-XL~\cite{wagner2020ptb} for in-domain evaluation and the Chapman-Shaoxing Database~\cite{zheng2022large} for cross-dataset evaluation. Experimental results show that DiSR-ECG achieves state-of-the-art SR performance on PTB-XL and demonstrates significantly stronger generalization to the Chapman-Shaoxing Database in a zero-shot setting. To the best of our knowledge, this is the first work to apply diffusion models to multi-lead ECG SR, bridging recent advances in generative modeling with the clinical demand for reliable and robust ECG reconstruction in real-world monitoring scenarios.

\section{Related Work}
\label{sec:rel}

\subsection{ECG super-resolution approach}
\label{ssec:ECGSR}

Conventional ECG SR methods, such as linear or cubic spline interpolation, distort the structure of ECG waveforms and may compromise the accuracy of diagnosis or applications. To address this issue, NNs have become mainstream in ECG SR by providing powerful nonlinear mapping capabilities. SRECG \cite{chen2023srecg} adapts SRResNet to ECG, using residual CNN blocks and deconvolution layers to upsample LR signals into HR outputs, and couples reconstruction with arrhythmia classification. DCAE-SR \cite{lomoio2024dcae} employs a denoising autoencoder with one encoder and two decoders: one branch reconstructs the noisy low-resolution signal, and the other generates a denoised high-resolution signal. 
MSECG \cite{lin2025msecg} integrates Mamba state-space models with lightweight convolutions, and it employs one-dimensional pixel-shuffle upsampling with a residual skip from the linearly upsampled input. These NN-based ECG SR techniques have been approached as discriminative methods, and their performance is only validated on in-domain scenarios.

\subsection{Diffusion models for super-resolution}
\label{ssec:diffusion}

Diffusion generative models have recently emerged as a powerful framework for SR in various domains, such as image and acoustic signals~\cite{yue2023resshift,9887996,han2022nu}. The core idea is to define a forward noising process that gradually corrupts the HR data $\bm{x}_0$ with Gaussian noise, and to learn a reverse process that restores the data distribution by iterative denoising.  

In the standard formulation, the forward process applies a variance schedule $\{\alpha_t\}_{t=1}^T$ that produces noisy samples $\bm{x}_t$:
\begin{equation}
q(\bm{x}_t \mid \bm{x}_0) = \mathcal{N}\!\big(\bm{x}_t;\, \sqrt{\bar{\alpha}_t}\bm{x}_0,\,(1-\bar{\alpha}_t)\bm{I}\big),
\end{equation}
where $\bar{\alpha}_t = \prod_{i=1}^t \alpha_i$. The reverse process is parameterized by a neural network $f_\theta$ trained to approximate the posterior distribution:
\begin{equation}
p_\theta(\bm{x}_{t-1}\mid \bm{x}_t,\bm{y}_0)
= \mathcal{N}\!\left(\bm{x}_{t-1}\,;\ \mu_\theta(\bm{x}_t,\bm{y}_0,t),\,\Sigma_t\right).
\end{equation}
which iteratively denoises $\bm{x}_t$ while conditioning on auxiliary inputs $\bm{y}_0$, such as the LR data.


\section{Proposed Method}
\label{sec:method}

\subsection{Residual shifting}
\label{ssec:resshift}

\begin{figure*}
    \centering
    \includegraphics[width=1\linewidth]{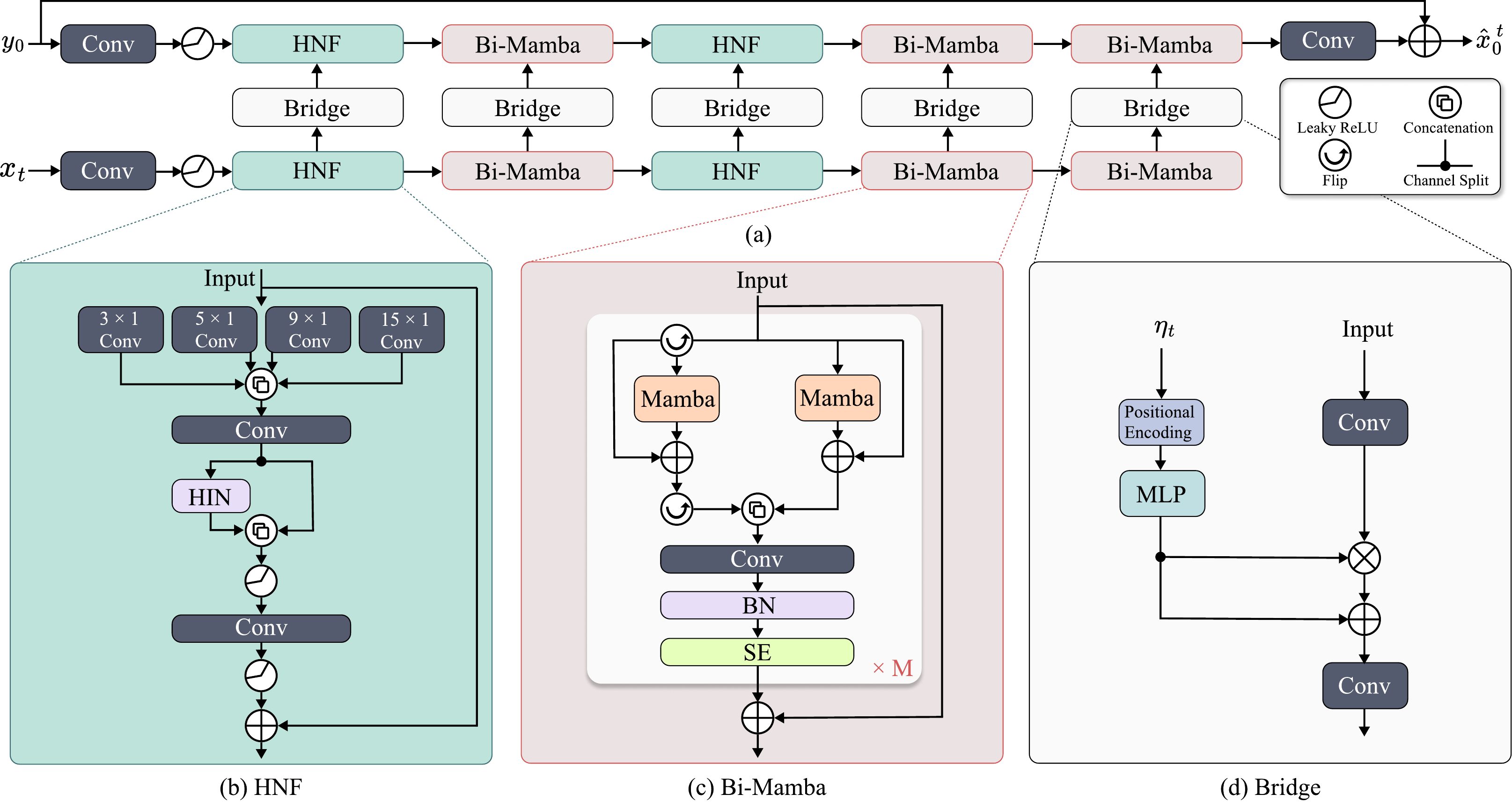}
    \caption{The model architecture of (a) DiSR-ECG, (b) HNF, (c) Bi-Mamba, and (d) Bridge.}
    \label{fig:architecture}
\end{figure*}

We propose an ECG SR diffusion framework based on Residual Shifting (ResShift)~\cite{yue2023resshift}, as shown in Fig.~\ref{fig:diffusionprocess}. 
ResShift redefines SR as a short Markov chain directly bridging the HR and LR domains. Instead of gradually diffusing the HR data into pure Gaussian noise, ResShift progressively shifts the residual between HR and LR while injecting noise at each step.
Consider a signal LR $\bm{y}_0$ and its corresponding HR signal $\bm{x}_0$. The residual between them is denoted as $\bm{e}_0 = \bm{y}_0 - \bm{x}_0$. A shifting sequence \(\{\eta_t\}_{t=1}^{T}\) follows a non-uniform geometric form~\cite{yue2023resshift}, with \(\eta_1 \to 0\) and \(\eta_T \to 1\). In the forward process, the resulting marginal distribution of $\bm{x}_t$ is given by:
\begin{equation}
\label{eq:forward}
q(\bm{x}_t \mid \bm{x}_0,\bm{y}_0)
= \mathcal{N}\!\Big(\bm{x}_t ; \bm{x}_0 + \eta_t\bm{e}_0,\; \kappa_t^2\,\eta_t\,\bm{I}\Big), 
\end{equation}
where $\kappa_t$ is the noise scale controlling the variance at step $t$.

In the reverse process, the corresponding reverse kernel is given by:
\begin{equation}
\label{eq:param-reverse}
p_\theta(\bm{x}_{t-1}\mid \bm{x}_t,\bm{y}_0)
= \mathcal{N}\!\Big(
     \bm{x}_{t-1}\;;
     \tfrac{\eta_{t-1}}{\eta_t}\bm{x}_t
     + \tfrac{\alpha_t}{\eta_t}\hat{\bm{x}}_0^{\,t},\;
     \kappa_t^{2}\tfrac{\eta_{t-1}}{\eta_t}\alpha_t\,\bm{I}
   \Big),
\end{equation}
where $\alpha_t = \eta_t - \eta_{t-1}$ for $t > 1$, $\alpha_1 = \eta_1$, and \(\hat{\bm{x}}_0^{\,t} = f_\theta(\bm{x}_t,\bm{y}_0,\eta_t)\) denotes the model’s estimate of the clean HR signal. \(f_\theta\) is a deep neural network parameterized by $\theta$, trained to predict $x_0$.
This design allows the sampling process to converge in fewer steps.

We modify the noise schedule \(\kappa\) for \(T\) steps from a fixed value ($\kappa=2$) to a linear form, so that the variance increases monotonically with \(t\).
The  \(\{\kappa_t\}_{t=1}^{T}\) is defined as:
\begin{equation}
\label{eq:kappa_linear}
\kappa_t = \kappa_{\min} + \tfrac{t}{T}\big(\kappa_{\max}-\kappa_{\min}\big), \qquad t=1,\ldots,T,
\end{equation}
where $\kappa_{\min} = 0.5$ and $\kappa_{\max} = 2.0$.
This preserves the signal structure in the early steps and gradually increases the stochasticity in the later steps.

For training, we compute reconstruction losses between the predicted $\hat{\bm{x}}_0^{\,t}$ and the ground truth HR signal $\bm{x}_0$. The primary objective is the $L_1$ reconstruction loss, which ensures accurate waveform recovery. 
To further emphasize signal fidelity, we add a scale-invariant signal-to-distortion ratio (SI-SDR) term \cite{LeRoux2018SDRH}, 
leading to the following objective:
\begin{equation}
\label{eq:loss_residual}
\mathcal{L}
=\big\|\hat{\bm{x}}_0^{\,t}-\bm{x}_0\big\|_{1}
+\lambda_{\mathrm{SI}}\!\left[-\,\mathrm{SI\!-\!SDR}\!\left(\hat{\bm{x}}_0^{\,t},\bm{x}_0\right)\right].
\end{equation}

At inference time, we employ the ensemble technique by drawing multiple stochastic samples from the trained model. Averaging these reconstructed waveforms reduces variance in high-frequency components and leads to more consistent ECG signals\cite{ensemble}.

\subsection{Network architecture}
\label{ssec:architecture}

Fig.~\ref {fig:architecture} (a) shows the model architecture in DiSR-ECG, which extends from DeScoD-ECG~\cite{li2023descod}. The network adopts a dual-stream design, where one stream extracts features from the noisy diffusion state \(\bm{x}_t\) 
and the other processes the low-resolution condition \(\bm{y}_0\). Each stream consists of two main modules: Half Normalized Filters (HNF)~\cite{li2023descod} and Bidirectional Mamba (Bi-Mamba) modules.
Feature interactions between the two streams are implemented through Bridge modules, which are inserted at multiple depths. In this way, condition information is progressively fused into the denoising pathway. After the last processing stage, the integrated features are projected onto 12 leads, yielding
\(f_\theta(\bm{x}_t,\bm{y}_0,\eta_t)\in\mathbb{R}^{12\times T}\). 
The final reconstruction is then obtained by residual addition with the condition input \(\bm{y}_0\), referred to as LR observation adding.

The HNF block, illustrated in Fig.~\ref{fig:architecture} (b), applies multi-scale convolutions to capture waveform patterns at different temporal resolutions, followed by channel aggregation. 
Half-instance normalization (HIN) is then applied, which has been shown to stabilize training while preserving the natural statistical properties of ECG signals~\cite{Chen_2021_CVPR}.

 Fig.~\ref{fig:architecture} (c) presents the structure of the Bi-Mamba module, which employs a bidirectional state space model to capture long-range temporal dependencies across beats and leads. Batch normalization (BN) is employed after convolution to improve training stability. To further refine the learned representations, a squeeze-and-excitation (SE)~\cite{Hu_2018_CVPR} gate is incorporated to adaptively reweight feature channels. The number of stacked Bi-Mamba layers is denoted by $M$ (set to $2$ in our experiments).

The Bridge block, shown in Fig.~\ref{fig:architecture} (d), performs cross-stream fusion. Following the feature-wise linear modulation (FiLM)~\cite{dumoulin2018feature}, each Bridge conditions on the diffusion scale \(\eta_t\) 
to modulate intermediate features.
Notably, we extend the FiLM mapping by employing a two-layer MLP (SiLU), 
which enables an affine \((\gamma,\beta)\) transformation.

\section{Experiments}

\subsection{Datasets}
\label{ssec:datasets}

This study employs ECG data from two open-access databases: PTB-XL~\cite{wagner2020ptb} and the Chapman-Shaoxing ECG Database~\cite{zheng2022large}. PTB-XL is a large clinical corpus containing 21,799 ten-second recordings at 500 Hz from 18,869 patients. Each ECG record is accompanied by metadata such as age, sex, and weight, along with manual annotations of signal quality. The Chapman-Shaoxing Database consists of 45,152 ten-second resting 12-lead ECGs at 500 Hz, with diagnostic labels covering a broad spectrum of cardiac abnormalities, including arrhythmias, conduction blocks, and myocardial infarction, making it both diverse and clinically representative. Both databases provide sufficient scale and diagnostic diversity for developing and evaluating ECG SR models.

To emulate real-world artifacts, we use the MIT-BIH Noise Stress Test Database (NSTDB)~\cite{moody1984bih}, which contains dedicated recordings of baseline wander (bw), muscle artifact (ma), and electrode motion (em). Each artifact type was acquired from volunteers over half-hour sessions using two channels at 360 Hz. These recordings are critical for stress-testing the robustness of ECG SR models.



\subsection{Data preprocessing and preparation}
\label{ssec:preprocess}

For training and in-domain evaluation, we use PTB-XL. For each 12-lead ECG record, we extract a 10\,s segment at 500\,Hz and apply a Butterworth bandpass filter (1-45\,Hz) for both LR ECG \(\bm{y}_0\) and HR ECG \(\bm{x}_0\). The LR ECG is formed by downsampling the ECG waveforms by a factor of 10 to 50\,Hz and then restoring the length (5{,}000 samples) by linear interpolation to 500\,Hz; the corresponding 500\,Hz ECG waveform forms the HR target \(\bm{x}_0\). We follow the official data splits for training (Folds 1-8), validation (Fold 9), and testing (Fold 10)~\cite{wagner2020ptb}.

Following prior works~\cite{lomoio2024dcae, lin2025msecg}, we stochastically add contaminants from the MIT-BIH NSTDB to generate noisy ECG data from PTB-XL. For each record, with probability 0.5, the sample remains clean; otherwise, we inject one contaminant type chosen from baseline wander (bw), electrode motion (em), or muscle activity (ma). We also sample a noise channel and a target Signal-to-Noise Ratio (SNR) from type-specific ranges~\cite{HU2024105504}. The randomly extracted noise segment is resampled to 50\,Hz, scaled to the target SNR by power matching on the LR ECG, and added per lead. 

For cross-dataset evaluation, we employ the Chapman-Shaoxing Database, applying the same preprocessing steps as PTB-XL but without noise injection.


\subsection{Evaluation metrics}
\label{sssec:metrics}

We evaluate the performance of ECG SR using four signal quality metrics following the prior work \cite{lin2025msecg}. 
Mean Squared Error (MSE) measures the average squared difference between the reconstructed signal and the ground truth, with lower values indicating closer waveform alignment. 
Cosine Similarity (CoS) quantifies the similarity in signal orientation, where values close to one reflect stronger morphological agreement between predicted and ground truth ECG. 
SNR measures the relative strength of the ground truth signal compared to the reconstruction error, with higher values corresponding to cleaner, more faithful signals. 
Maximum Absolute Deviation (MAD) captures the largest pointwise error across the segment, where lower values denote better reconstruction with smaller outliers. 

\subsection{Implementation details}
\label{ssec:implementation}
In the diffusion setting, the shift schedule \(\{\eta_t\}_{t=1}^{T}\) is geometric with exponent \(p=0.3\), and the number of steps is set to 15.  In training, we optimize with Adam~\cite{kingma2014adam} at an initial learning rate of \(1\times10^{-4}\), which is reduced by a factor of \(0.1\) at epoch \(150\). We train for \(400\) epochs with batch size \(16\), apply gradient norm clipping at \(1.0\), and maintain an exponential moving average of the parameters with decay \(0.99\) throughout training. The SI-SDR loss is weighted by \(\lambda_{\text{SI}}=0.02\). For inference, we employ 10 samples for the ensemble method.

\subsection{Results and discussion}
\label{ssec:result}
\begin{table}[t]
\centering
\caption{Overall performance comparison of ECG SR methods on the PTB-XL dataset (in-domain evaluation).}
\scriptsize
\begin{tabular}{lcccc}
\toprule
Method &  MSE($\times 10^{-3}$)$\downarrow$  & 
CoS ($\times 10^{-2}$)$\uparrow$ & SNR (dB)$\uparrow$ & MAD$\downarrow$ \\
\midrule
LI       & 7.477  & 90.211 &  8.592  & 0.877 \\
SRECG~\cite{chen2023srecg}  & 0.422 & 99.367 & 19.751 & 0.371 \\
DCAE-SR~\cite{lomoio2024dcae} & 4.461 & 97.560 & 12.591  & 0.962  \\
MSECG~\cite{lin2025msecg} & 0.184 & 99.745 & 24.037 & 0.221 \\
\midrule
DiSR-ECG & \textbf{0.181} & \textbf{99.754}$^{*}$  & \textbf{24.092}$^{*}$ & \textbf{0.216} \\
\bottomrule
\multicolumn{5}{l}{* denotes statistical significance over MSECG $(p<0.05)$.} \\
\end{tabular}
\label{tab:overall}
\end{table}

\begin{table}[t]
\centering
\caption{Cross-dataset evaluation results on the Chapman–Shaoxing Database in a zero-shot setting.}
\scriptsize
\begin{tabular}{lcccc}
\toprule
Method &  MSE ($\times 10^{-3}$)$\downarrow$  & 
CoS ($\times 10^{-2}$)$\uparrow$ & SNR (dB)$\uparrow$ & MAD$\downarrow$ \\
\midrule
LI       & 4.289 & 95.114 & 10.549 & 1.111 \\
SRECG~\cite{chen2023srecg}  & 1.227 & 98.729 & 17.535 & 0.659 \\
DCAE-SR~\cite{lomoio2024dcae} & 9.017 & 96.168 & 10.829 & 1.410 \\
MSECG~\cite{lin2025msecg} & 0.776 & 99.084 & 21.332 & 0.487 \\
\midrule
DiSR-ECG & \textbf{0.642}$^{*}$ & \textbf{99.376}$^{*}$ & \textbf{21.939}$^{*}$ & \textbf{0.447}$^{*}$ \\
\bottomrule
\multicolumn{5}{l}{* denotes statistical significance over MSECG $(p<0.05)$.} \\
\end{tabular}
\label{tab:ecg}
\end{table}

\begin{figure}[t]
\centering
\includegraphics[width=1\linewidth]{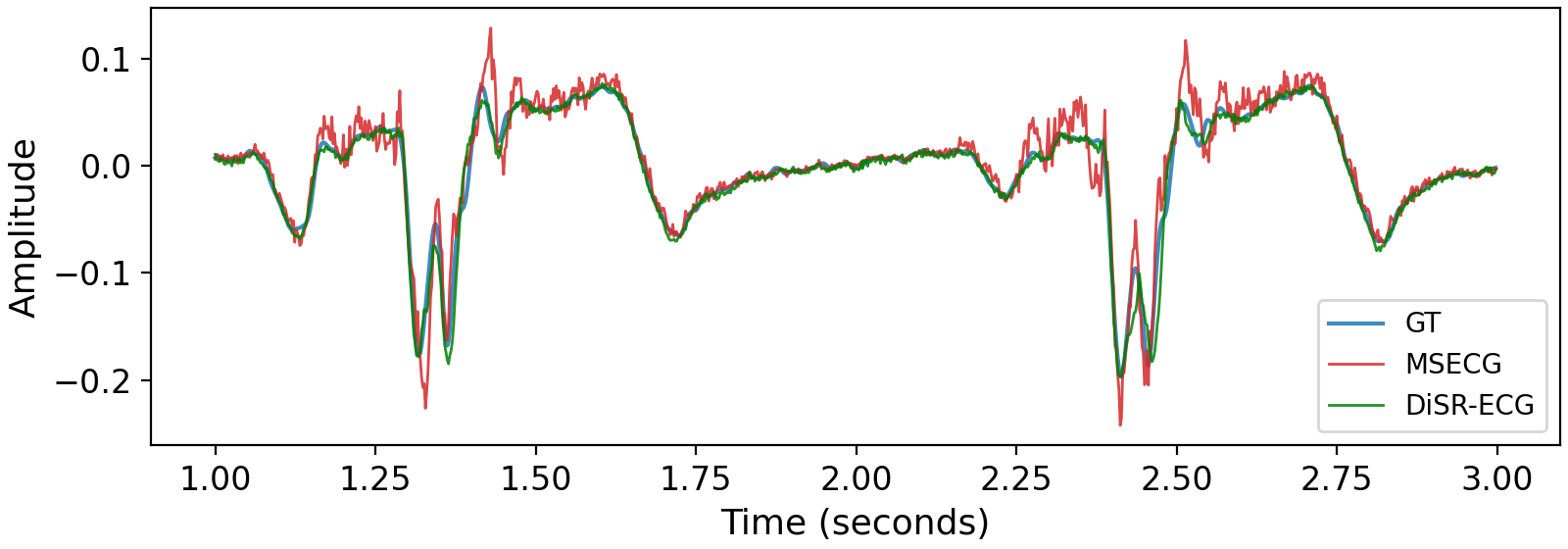}
\caption{Qualitative comparison of reconstructed waveforms on the Chapman-Shaoxing Database.}
\label{fig:qual}
\end{figure}

\begin{table}[t]
\centering
\caption{Ablation study evaluating the effects of the loss function, noise schedule, and LR observation adding.}
\setlength{\tabcolsep}{4pt}
\scriptsize
\begin{tabular}{lcccc}
\toprule
Method & MSE ($\times 10^{-3}$) $\downarrow$ & CoS ($\times 10^{-2}$) $\uparrow$ & SNR (dB) $\uparrow$ & MAD $\downarrow$\\
\midrule
DiSR-ECG      & \underline{0.181} & \textbf{99.754} & \textbf{24.092} & \underline{0.216} \\
\midrule
w/o SI-SDR loss          &\textbf{0.180}& 99.735 & 23.696 &   0.233 \\

$\kappa_t$ fixed      & 0.213 & 99.691 & 22.882 & \textbf{0.212} \\

w/o LR adding    & 0.192 &  \underline{99.747} &  \underline{23.824} & 0.220 \\

\bottomrule
\multicolumn{5}{l}{\textbf{Bold} denotes the best performance; \underline{underline} denotes the second best performance.}
\end{tabular}
\label{tab:ablation}
\end{table}

We compare DiSR-ECG with a conventional SR method (LI) and three NN-based approaches (SRECG, DCAE-SR, and MSECG). The overall performance on PTB-XL is summarized in Table~\ref{tab:overall}. DiSR-ECG achieves the best results across all four quality metrics. Notably, compared with MSECG, which already demonstrates strong in-domain performance, DiSR-ECG still provides statistically significant improvements in SNR and CoS ($p<0.05$, paired-sample t-test).

Table~\ref{tab:ecg} reports the cross-dataset results evaluated in a zero-shot setting on the Chapman-Shaoxing Database. As expected, the performance of all methods degrades under distribution shift. However, the proposed method consistently outperforms all baselines. In particular, compared with the strongest baseline MSECG, DiSR-ECG exhibits less performance degradation, achieving higher scores in all four metrics with statistical significance ($p<0.05$). This finding confirms that DiSR-ECG provides more reliable generalization under domain mismatch than prior discriminative NN-based methods.

Fig.~\ref{fig:qual} presents reconstructed ECG waveforms from the Chapman-Shaoxing Database (record JS00025, lead aVR, annotated as LBBB). The waveform produced by DiSR-ECG aligns more closely with the ground-truth (GT) signal than that of MSECG, which exhibits random high-frequency jitter around the QRS complex and severe structural distortion. These results further confirm the robustness of DiSR-ECG under domain-shift conditions.

Table~\ref{tab:ablation} presents an ablation study examining the contribution of each component in DiSR-ECG using PTB-XL. Training without SI-SDR loss slightly reduces reconstruction quality: while MSE decreases marginally, SNR drops, and MAD increases, indicating that SI-SDR helps maintain a more stable waveform structure. Replacing the linear noise schedule with a fixed $\kappa_t$ degrades three metrics (MSE, CoS, and SNR), confirming that the adaptive schedule is crucial for balancing detail preservation and stochasticity during sampling. Removing the LR observation adding technique also leads to moderate decreases, highlighting its role in stabilizing reconstruction. Collectively, these results demonstrate that each component of DiSR-ECG plays an important role in the final performance.

\section{Conclusion}
\label{sec:conclusion}

In this study, we introduced DiSR-ECG, a residual shifting conditional diffusion framework for ECG super-resolution. Experimental results showed that DiSR-ECG outperforms baseline methods on the large-scale PTB-XL database and generalizes more reliably to the Chapman-Shaoxing Database under distribution shifts. These findings highlight the potential of diffusion-based SR for producing high-fidelity ECG reconstructions that could support downstream tasks such as arrhythmia diagnosis and rhythm analysis. Future work will focus on optimizing inference efficiency for real-time deployment through lightweight architectures or flow-matching techniques. We also plan to evaluate DiSR-ECG on downstream tasks, including ECG delineation and arrhythmia detection, to further validate its clinical utility.

\bibliographystyle{IEEEtran}
\bibliography{refs_1}
\end{document}